\documentclass[11pt]{statsoc}
\usepackage{amsmath,natbib}
\usepackage{bm} 
\usepackage{graphicx,psfrag}
\usepackage{epstopdf}
\usepackage{verbatim}
\usepackage{upref}
\usepackage{amssymb, eucal} 
\usepackage{color}

\newtheorem{theorem}{Theorem}

\newcommand*{\bs}{\boldsymbol}

\title[Density Estimation Using Transformations]{Two Dimensional Density Estimation using Smooth\\ Invertible Transformations} 
\author{Ethan Anderes} 
\address{University of California at Berkeley, USA.} 
\email{anderes@stat.berkeley.edu} 
\author[E. Anderes and M. Coram]{Marc Coram} 
\address{Stanford University, USA.} 
\email{mcoram@stanford.edu}

\begin{document}

\begin{abstract} 
We investigate the problem of estimating a smooth invertible transformation $f$ when observing independent samples $X_1,\ldots,X_n\sim \Bbb P\circ f$ where $\Bbb P$ is a known measure. We focus on the two dimensional case where $\Bbb P$ and $f$ are defined on $\Bbb R^2$. We present a  flexible class of smooth invertible transformations in two dimensions with variational equations for optimizing  over the classes, then study the problem of estimating the transformation $f$  by penalized maximum likelihood estimation.  We apply our methodology to the case when $\Bbb P\circ f$ has a density with respect to Lebesgue measure on $\Bbb R^2$ and demonstrate improvements over kernel density estimation on three examples.
\end{abstract}



\section{Introduction}
In this paper we investigate the problem of estimating the probability distribution of a random vector $X$ 
from independent copies,
\[ X_1,\ldots,X_n\overset{iid}\sim \Bbb P\circ f,\] where $\Bbb P$ is a known measure on $\Bbb R^d$ and $f$ is an unknown smooth bijection of $\Bbb R^d$.  We specifically focus our attention to the two dimensional case, where the theory of quasiconformal maps is at our disposal.  
The estimate is constructed based on the fact that the distribution of $X$ is characterized by $f(X)\sim \Bbb P$. Therefore,  by attempting to ``deform'' the data $X_1,\ldots,X_n$ by a transformation $f$ which satisfies, 
\[ f(X_1),\ldots,f(X_n)\overset{iid}\sim \Bbb P, \]
we get an estimate of the distribution of $X$. In what follows we iteratively  deform the data toward the target distribution $\Bbb P$ using a gradient ascent type algorithm  over a  class of transformations. 
 For the remainder of this paper we reserve the term \lq transformation'  or  \lq deformation' to refer to a smooth map which is a bijection of $\Bbb R^d$, i.e. invertible and surjective. 

 Deformations have been used in many types of statistical problems. For example,  deformations are used in spatial statistics for developing nonstationary random fields (see \cite{SamGu:nonstat},   \cite{SchmidtOHagan:def}, \cite{DamianSampsonGuttorp:def},  \cite{PerrinMeiring:def1999},  \cite{IovleffPerrin:Def2004},  \cite{ClercMallat:def}, \cite{AnderesStein:paper}). Another example is the use of
deformable templates for  computer vision and medical imaging problems (see \cite{Younes1999}, \cite{Joshi2000}, \cite{Dupuis1998}, \cite{BajcsyBroit1982}, \cite{Bajcsy1983}, \cite{Bookstein1989},  \cite{Amit1994}).
There has also been some work on using transformations to boost the efficiency of kernel density estimates.
 For example, in the one dimensional setting  \cite{ruppert1994transdensity} estimate a transformation of data to a uniform distribution in order to reduce the bias of kernel density estimation.  
In a similar manner, data sharpening techniques,  developed by \cite{choi1999sharp} and \cite{hall2002sharp}  are used for improving linear density estimates and are available in multiple dimensions. These techniques perturb the data slightly to produce bias reduction. However,  the perturbations need not be bijective, and so do not represent proper transformations.

Although deformations provide a powerful modeling tool, there are many challenges with their practical implementation.  The invertability condition, in particular, is one of the main difficulties for constructing flexible classes of transformations and searching within these classes. One of the most common techniques for dealing with the invertability condition is the use of time varying vector field flows for constructing deformations.
 These  vector fields $\{ u_t\}_{t\geq 0}$ define deformations $\{f_t\}_{t\geq 0}$ indexed by time $t\geq0$ which can be used for finding a minimizer of some data dependent penalty.  In this paper we use a different characterization of deformations to construct classes of deformations  indexed by a parameter vector $\bs \theta$. However, we derive a correspondence between our parameterization and a time varying vector field characterization of the corresponding deformations, whereby relating the two characterizations.       
     
Our basic approach for estimating $f$ is to maximize a penalized likelihood over classes of smooth transformations. Of course, this requires the existence of a density, which will be guaranteed by mild smoothness conditions on $\Bbb P$ and $f$. In two dimensions, we develop flexible classes of smooth invertible transformations generated by basis functions,  then use the theory of quasiconformal maps to derive variational formulas that relate perturbations of a parameterization to perturbations of the transformations. 
This, along with a formula for the rate of change of the likelihood allows us to search the parameter space using gradient-based optimization of a penalized likelihood. 

We finish this section with a more detailed overview of the results of this paper which, we hope, will help the reader follow the remainder of the paper. In section \ref{class}, we construct our class of deformations  by using a nonlinear transform of the linear span of a set of basis functions, the coefficients of which generate the parameterization $\mathcal F=\{f^{\bs \theta}\colon \bs\theta\in\Theta\subset \Bbb R^\infty  \}$.   We then use known results from quasiconformal maps, outlined in Section \ref{quasi22},  to  derive variational relationships that relate a perturbation of the parameters $\bs\theta$ to the rate of change of the likelihood $\ell(\bs \theta)$ (see Sections \ref{class} and \ref{RateLike}). In particular, let $\bs \theta$ and $d\bs \theta$ be two parameter vectors such that $\bs \theta+\epsilon d\bs \theta\in\Theta$ for all sufficiently small $\epsilon>0$.  The results in Section \ref{Vary1} show that
 \begin{equation} 
 \label{firsteq}
 f^{\bs \theta + \epsilon d\bs\theta}= f^{\bs \theta}+\epsilon u^{\bs\theta,d\bs\theta}\circ f^{\bs \theta} +o(\epsilon) 
 \end{equation}
as $\epsilon\rightarrow 0$, where $u^{\bs \theta,d\bs\theta}$ is a vector field  which depends on both $\bs \theta$ and $d\bs\theta$.   In Section \ref{algo}, we present some of the algorithmic details for numerically recovering $u^{\bs\theta,d\bs\theta}$ from $\bs \theta$ and $d\bs \theta$. We notice that under mild conditions there exists a partial differential equation which  is numerically solved to approximate $u^{\bs \theta,d\bs\theta}$. 

The variational results for $f^{\bs \theta}$ are developed primarily to allow one to perform a gradient-based optimization of a penalized version of the  likelihood of $X_1,\ldots,X_n\overset{iid}\sim \Bbb P\circ f^{\bs \theta}$. For the existence of a likelihood one needs sufficient smoothness conditions to ensure $\Bbb P\circ f^{\bs \theta}$ has a density. To this end, we assume that $\Bbb P$ has  a strictly positive differentiable density with respect to Lebesgue measure on $\Bbb R^2$. By writing the density as $\exp H$ for some  differentiable function $H\colon\mathbb R^2\rightarrow \mathbb R$ and by assuming $f$ is an orientation preserving $C^1$ diffeomorphism of $\Bbb R^2$,
 $\Bbb P\circ f$ has density $ |J_f|  \exp H\circ f$ where $|J_f|$ is the  determinant of the Jacobian of $f$ (always positive by the orientation preserving assumption on $f$).
 Let  $\ell(\bs\theta)$ denote the log likelihood of the sample $X_1\ldots,X_n\overset{iid}\sim \Bbb P\circ f^{\bs \theta}$ so that
\begin{equation}
\label{loglike}
\ell(\bs\theta)=\sum_{k=1}^n\log |J_f^{\bs \theta}(X_k)| +H\circ f^{\bs \theta}(X_k).
\end{equation}
 Now define the rate of change of the log likelihood at $\bs \theta$ in the direction $d\bs\theta$ by
$\dot\ell[\bs \theta](d\bs \theta):=\lim_{\epsilon\rightarrow 0}\frac{\ell(\bs \theta+\epsilon d\bs \theta)-\ell(\bs \theta)}{\epsilon}$ when it exists. In Section \ref{RateLike} we derive the following formula for $\dot\ell$
\begin{equation} 
\label{like}
\dot\ell[\bs \theta](d\bs\theta)=\sum_{k=1}^n \text{div}\, u^{\bs\theta,d\bs\theta}(Y_k) +\langle u^{\bs\theta,d\bs\theta}(Y_k),\nabla H(Y_k) \rangle
\end{equation}
where $Y_k=f^{\bs \theta}(X_k)$, $\langle \cdot,\cdot \rangle$ is dot product in Euclidean space, $u^{\bs\theta,d\bs\theta}$ is the vector field  found in (\ref{firsteq}), and $\text{div}\,u^{\bs\theta,d\bs\theta}$ is the divergence of $u^{\bs\theta,d\bs\theta}$. A heuristic interpretation of (\ref{like}) can be given by viewing the two terms as being competitors for maximizing the rate of increasing of the log-likelihood. In particular, let $f$ be a proposed transformation of the data so that $f(X_1),\ldots,f(X_n)$ are approximately distributed as $\Bbb P$. Consider attempting a small perturbation by a vector field $u$, $f+\epsilon u\circ f+o(\epsilon)$, that increases the log-likelihood. To find a good perturbation one wants to maximize the two terms: $\text{div}\,u( f(X_k))$ and $\langle u(f(X_k)),\nabla H(f(X_k)) \rangle$, for $k=1\ldots, n$. Notice that  $\text{div}\,u( f(X_k))$ measures local expansion at $f(X_k)$ so that large values of  $\text{div}\,u( f(X_k))$ correpond to an expansion the region surrounding  $f(X_k)$. In contrast, the term $\langle u(f(X_k)),\nabla H(f(X_k)) \rangle$  is large when the vector field gravitates toward the  modes of the density of $\Bbb P$. This gives an expansion-contraction competition for increasing the  rate of change of the log-likelihood and by suitably balancing these two competing terms one gets the best rate of increase of the log-likelihood.

In Section \ref{algo} we  use equations (\ref{firsteq}) and (\ref{like}) to perform a gradient ascent of a penalized version of the log likelihood.
By truncating the parameter space $\bs \theta^N:=(\theta_1,\ldots,\theta_N)$,  the gradient ascent path $\{\bs \theta^N_t \}_{t\geq 0}$ is characterized by
\[  \frac{d}{d t} \bs\theta^N_t =\nabla \ell (\bs \theta_t^N)- \lambda \nabla \mathcal J(\bs \theta_t^N)\]
where $\mathcal J$ is a regularization penalty, $\lambda\geq 0$ is a tuning parameter  and
\[ \nabla \ell (\bs \theta^N)=  \begin{pmatrix} \dot{\ell}[\bs\theta^N_t](e_1)\\\vdots\\ \dot{\ell}[\bs\theta^N_t](e_N) \end{pmatrix}, \]
where $e_1,\ldots,e_N$ are the standard basis vectors for $\Bbb R^N$.
The numerical details for computing $\nabla \ell (\bs \theta^N)$ and implementing the gradient-based optimization are also given in Section \ref{algo}.
Finally, in Section \ref{empir}, we present a series of simulation experiments. Our estimates, which are constructed using a Fourier basis to generate the parameterization $f^{\bs\theta}$,  compare favorably to a kernel density estimate.
\section{Quasiconformal maps and Variations}
\label{quasi22}
We briefly review some results from the theory of quasiconformal maps.  For more a complete treatment see \cite{ahlfors2}, \cite{krushkal:book}, \cite{law:book}, \cite{lehto}.
The main result of quasiconformal theory used in this paper is the characterization of quasiconformal maps by another function called a complex dilatation. Complex dilatations are important  for two reasons. First, they are only required to be measurable and bounded above by some $k<1$, which makes it easy to construct classes of smooth transformations.  Secondly, there is a constructible correspondence between the perturbation of a complex dilatation and the  perturbation of the corresponding transformation. This is used to derive the rate of change of the likelihood (see equation (\ref{like})) as one varies a parameterization of smooth transformations.

For a $C^1$ diffeomorphism $f$ define
\[ \partial f:=\frac{1}{2}\left(\frac{\partial }{\partial x}-i\frac{\partial}{\partial y}\right)f,\qquad \overline \partial f:= \frac{1}{2}\left(\frac{\partial }{\partial x}+i\frac{\partial}{\partial y}\right)f. \]
The complex dilatation of $f$, denoted $\mu_f$ or just $\mu$, is defined as $\mu:=\overline \partial f/\partial f$.  The  value of the complex dilatation at a point $z$, $\mu(z)$, characterizes the infinitesimal ellipse about $z$, which gets mapped to a infinitesimal circle under the image of $f$. In particular, the eccentricity of the ellipse is given by $\frac{1+|\mu|}{1-|\mu|}$  and the inclination is given by $\arg(-\mu/2)$. We will sometimes use the alternative notation $f_z$, $f_{\overline{z}}$ for $\partial f$, $\overline{\partial }f $ when doing so makes an equation more readable.

By generalizing to weak  derivatives for $\partial f$ and $\overline \partial f$, one can extend the definition of the complex dilatation $\mu_f:=\overline \partial f/\partial f$.  A quasiconformal map is defined to be a  homeomorphism $f$ with locally square integrable weak derivatives such that $\text{ess}\sup |\mu_f|<1$.  This condition on $\mu_f$ is equivalent to requiring that the eccentricities of the local ellipses characterized by $\mu_f$ be a.e. bounded. We now have the following Theorem found in \cite{ahlfors2}, page 57.

\begin{theorem} \label{mappingthm} For any measurable $\mu$ with $\|  \mu\|_\infty<1$ there exists a unique quasiconformal mapping $f^\mu$ with complex dilatation $\mu$ that leaves  $0,1,\infty$ fixed.
\end{theorem} 
We also have the nice fact that all quasiconformal maps with a given dilatation are found by post composing the unique map $f^\mu$ in Theorem \ref{mappingthm} with a conformal map. In this way, any quasiconformal map  $f\colon\Bbb C\rightarrow \Bbb C$ with dilatation $\mu$ has the form $a f^\mu +b$ for $a, b\in\Bbb C$.
The usefulness of this characterization is seen in the  construction of classes of quasiconformal maps in Section \ref{class}.
We finally mention that if $\mu$ is known to be H\"older continuous, the map $f$ will be a $C^1$ diffeomorphism.

\subsection{The dependence of $f^\mu$ on $\mu$} 
Besides characterizing quasiconformal maps, the other important property of the complex dilatation   is that  $f^\mu$ depends differentially on $\mu$ so that a small perturbation $\mu+\epsilon \nu$  leads to a vector field perturbation $f^\mu +\epsilon u\circ f^\mu + o(\epsilon)$ of $f^\mu$. An integral equation then relates  $\nu$ and the vector field $u$. 
 
 To be more precise suppose $\{ \mu_t \}_{t\geq 0}$ is a class of deformations such that $\mu_{t+\epsilon}=\mu_t+\epsilon \nu_t+\epsilon \gamma(t,\epsilon)$ where $\gamma(t,\epsilon),\nu_t\in L_{\infty}$, $\| \mu_t \|_\infty<1$, and $\| \gamma(t,\epsilon) \|_\infty\rightarrow 0$ as $\epsilon\rightarrow 0$. Then by Theorem 5, page 61 of \cite{ahlfors2},
 \begin{equation}\label{perturb} f^{\mu_{t+\epsilon}}=f^{\mu_t}+\epsilon u_t\circ f^{\mu_t}+o(\epsilon),\end{equation}
 uniformly on compact subsets where,
 \begin{equation} \label{cauchyintegral} u_t(\zeta):=-\frac{1}{\pi}\iint {L^{\mu_t}\nu_t}(z)\,R(z,\zeta) dxdy, \end{equation}
 for  $R(z,\zeta)=\frac{\zeta(\zeta-1)}{z(z-1)(z-\zeta)}$, $  L^{\mu} \nu:=\left\{ \frac{\nu}{1-|\mu|^2}\frac{\partial f^{\mu}}{\overline{\partial f^{\mu}}} \right\}\circ (f^\mu)^{-1}$ and $z=x+iy$. By H\"older's inequality the integral indeed  exists when $L^{\mu_t}\nu_t\in L_\infty$. By properties of the integral (\ref{cauchyintegral}) the vector field $u_t$ satisfies 
 \begin{equation} 
 \label{diffeq1}
  \overline\partial u_t=L^{\mu_t}\nu_t,\quad u_t(0)=u_t(1)=0,\end{equation}
 in the distributional sense when $L^{\mu_t}\nu_t\in L_p$ for some $p>2$ (see Lemma 3, page 53 of \cite{ahlfors2}).  
 We use (\ref{diffeq1})  to approximate the integral  (\ref{cauchyintegral}) in Section \ref{algo}.
 
 Finally, one of the consequences  of (\ref{perturb}) is that for sufficiently smooth $\{ u_t\}_{t\geq 0}$  the determinant of the Jacobian of $f^{\mu_t}$ at $x\in \Bbb R^2$, denoted $|J_t(x)|$, exists and satisfies 
 \begin{equation} 
 \label{jacobiId}
 \frac{d}{dt} |J_{t}(x)|= |J_t(x)| \text{div}\, u_t\circ f_t(x) .\end{equation} 
For references see  Theorem 3.1.2 of \cite{Hille}  or Corollary 3.1 of \cite{hartmanBook} (page 96).
 It is  (\ref{jacobiId}) that gives the rate of change of the likelihood $\dot\ell[\bs \theta](d\bs\theta)$ in equation (\ref{like}).
 %

\section{Classes of Quasiconformal maps}
\label{class}
To ensure that $\Bbb P\circ f$ is a genuine probability measure, one must require that the $C^1$ diffeomorphism, $f$, be invertible.
It is this nonlinear  condition  that is the main obstacle for constructing a rich class of transformations, $\{ f^{\bs \theta}\colon \bs\theta\in\Theta\subset \Bbb R^\infty \}$. In this section, to overcome this difficulty, we use the flexibility of complex dilatations $\mu$ to construct our class of transformations.

By defining a  set of basis functions $\varphi_k:\mathbb C\rightarrow \mathbb C$ for $k=1,2,\ldots$ we formally construct a class of complex dilatations,
\begin{equation}
\label{classes}
 \mathcal M:=\left\{\mu=\frac{\varphi}{1+|\varphi|}: \varphi=\sum_k c_k \varphi_k,c_k\in\Bbb R   \right\}. 
 \end{equation}
The reason for introducing the transformation $\varphi\mapsto \frac{\varphi}{1+|\varphi|}$ is to remove any restriction, besides measurability and boundedness, on the linear span $\varphi=\sum_k\varphi_k$ to define a class of quasiconformal maps. 
Indeed, by Theorem \ref{mappingthm}, for any complex dilatation $\mu\in \mathcal M$ with $\varphi=\sum_{k} c_k \varphi_k\in L_\infty$, there exists a unique quasiconformal map $f^{\bs \theta}=a f^\mu +b$, where $\bs \theta=(a_1,a_2,b_1,b_2,c_1,c_2,\ldots)$, $a=a_1+ia_2\neq 0$ and $b=b_1+ib_2$.
Now we construct the full class $\mathcal F$ of transformations,
 \[\label{classesofdef} \mathcal F:=\{ f^{\bs \theta}\colon \bs\theta\in\Theta\subset \Bbb R^\infty \},\]
where $\Theta$ is the set of parameter values for which $\sum_{k} c_k \varphi_k\in L_\infty$ and $a\neq 0$.

 \subsection{Variational Relationship between $\bs \theta$ and $f^{\bs \theta}$} 
 \label{Vary1}
One of the main components of the gradient ascent algorithm is the computation of the vector field perturbation of $f^{\bs \theta}$ that results from a small change of the parameter vector $\bs \theta+\epsilon d\bs \theta$. We start this section by studying how the perturbation $\bs \theta+\epsilon d\bs \theta$ propagates to  $\mu$, then finish with a derivation of the expansion $f^{\bs \theta}+\epsilon u^{\bs \theta,d\bs\theta}\circ f^{\bs \theta}+o(\epsilon)$.

Suppose $\varphi,d \varphi\in L_\infty$ are both members of the linear span of $\{\varphi_k \}_{k\in \Bbb N}$.
We start by showing that perturbing $\varphi$ by 
 $\varphi+\epsilon d\varphi$ results in  $ \mu  +\epsilon \nu+o(\epsilon)$ where
\begin{equation}\label{llastequn} 
\nu=
d\varphi\, \frac{2+|\varphi| }{2(1+|\varphi|)^2} -  \overline {d\varphi} \,\frac{ \varphi^2}{2|\varphi|(1+|\varphi|)^2}  \end{equation}
where, for the rest of this paper,  the second term of the right hand side of (\ref{llastequn}) is understood to be zero at the points $z$ such that $\varphi(z)=0$. 
To be more precise, let $\mu_\varphi=\frac{\varphi}{1+|\varphi|}$ and $\nu_\varphi(d\varphi):=\nu$ as in (\ref{llastequn}).
Suppose $\varphi=\sum_{k=1}^\infty c_k$ and $d\varphi=\sum_{k=1}^\infty dc_k\, \varphi_k$  are complex functions in $L_\infty$ with real coefficients $\{ c_k\}_{k\geq 1}$ and $\{ dc_k \}_{k\geq 1}$. Then 
\begin{equation}
\label{showme4}
\mu_{\varphi+\epsilon d\varphi}=\mu_\varphi+\epsilon \nu_\varphi (d\varphi) + \epsilon \gamma (\epsilon)
\end{equation}
where $\nu_\varphi(d\varphi),\gamma(\epsilon)\in L_\infty$ and $\| \gamma(\epsilon) \|_\infty\rightarrow 0$ as $\epsilon \rightarrow 0$.
To see why, let $\mathcal T(z)=\frac{z}{1+|z|}$, where $z=x+iy$ and notice that $\partial \mathcal T(z)/\partial x$ and $\partial \mathcal T(z)/\partial y$ are continuous on $\Bbb C$ and satisfy
\begin{align*}
\frac{\partial \mathcal T(z)}{\partial x}&=\frac{1}{1+|z|}-x\frac{z}{|z|(1+|z|)^2} \\
\frac{\partial \mathcal T(z)}{\partial y}&=\frac{i}{1+|z|}-y\frac{z}{|z|(1+|z|)^2}.
\end{align*}
where the last term is $0$ when $z=0$.  
Therefore directional derivative $\partial _u \mathcal T$, in the direction $u\in\Bbb C$, exists and is continuous on $\Bbb C$ for any fixed $u=u_1+iu_2$ and is given by 
\begin{align*} 
\partial_u\mathcal T(z)&= u_1\frac{\partial \mathcal T(z)}{\partial x}+ u_2 \frac{\partial \mathcal T(z)}{\partial y} \\
&=\frac{u}{1+|z|} - \left(\frac{u\overline z+\overline u z}{2}  \right)\left(\frac{z}{|z|(1+|z|)^2}  \right)\\
&=u\frac{2+|z|}{2(1+|z|)^2} - \overline u \frac{z^2}{2|z|(1+|z|)^2} \\
\end{align*}
where the last term is $0$ when $z=0$.  Therefore the map $\mathcal T$ has a Fr\'echet derivative $\mathcal T^\prime_z(u):=\partial_u \mathcal T(z)$ and by the mean value theorem (see \cite{Dieudonne})
\[  \left| \frac{\mathcal T(\varphi(z)+\epsilon d\varphi(z))- \mathcal T(\varphi(z))}{\epsilon} - \mathcal T_{\varphi(z)}^\prime(d\varphi)\right|\leq |d\varphi(z)|\sup_{\xi \in S} \left \| \mathcal T^\prime_\xi -\mathcal T^\prime_{\varphi(z)} \right\|\]
for a fixed $z\in \Bbb C$, where $S$ is the line connecting $\varphi(z)+\epsilon d\varphi(z)$ and $\varphi(z)$, and $\| \cdot \|$ is the standard operator norm on the space of continuous linear mappings from $\Bbb R^2$ to $\Bbb R^2$. Now
\begin{align*}
\sup_{\xi \in S}\left\| \mathcal T^\prime_\xi -\mathcal T^\prime_{\varphi(z)}  \right\| &\leq \sup_{\small|\xi-\varphi(z)|\leq \epsilon B}\left\| \mathcal T^\prime_\xi -\mathcal T^\prime_{\varphi(z) } \right\|  
\end{align*}
where $\| d\varphi \|_\infty=B$. The last term converges to zero uniformly in $z\in\Bbb C$ as $\epsilon\rightarrow 0$ since $\varphi(z)$ is bounded and $\mathcal T^\prime_{\eta} $ is uniformly continuous on compact domains. Therefore 
\[ \mathcal T(\varphi +\epsilon d\varphi)=\mathcal T(\varphi)+\epsilon \mathcal T_\varphi^\prime(d\varphi)+\epsilon \gamma(\epsilon) \]
where  $\|\gamma (\epsilon) \|_\infty\rightarrow 0$
as $\epsilon\rightarrow 0$, which gives (\ref{showme4}).
Finally  notice $|\nu_\varphi (d\varphi)|=|\mathcal T_\varphi^\prime(d\varphi)|\leq \| d\varphi \|_\infty/(1+|\varphi|)\leq \|d\varphi \|_\infty<\infty$. 

Now we can combine all the perturbation results to derive variational relationship between  the coefficients and the maps.   Suppose $\bs \theta=(\theta_1,\theta_2,\ldots)$ and $d\bs\theta=(d\theta_1,d\theta_2,\ldots)$ are two parameter vectors such that $\bs\theta+\epsilon d\bs\theta\in \Theta$ for all sufficiently small $\epsilon>0$. Let $\varphi=\sum_{k=4}^\infty \theta_k\varphi_k$, $d\varphi=\sum_{k=4}^\infty d\theta_k \varphi_k$, $\mu=\frac{\varphi}{1+|\varphi|}$, and $\nu$ defined by (\ref{llastequn}).  Also let  $da=d\theta_1+id\theta_2$ and similarly $db=d\theta_3+id\theta_4$.
The following display summarizes how the perturbation of $\bs \theta$ by $\epsilon d\bs\theta$ propagates through $f^{\bs\theta}$:
\begin{align*}
\bs\theta+\epsilon d\bs \theta &\mapsto 
\left[ \begin{array}{c} \varphi+\epsilon d\varphi \\
a+\epsilon da \\
b+\epsilon db
\end{array}\right] 
\overset{\text{by (\ref{showme4})}}
\mapsto 
\left[ \begin{array}{c} \mu+\epsilon \nu +o(\epsilon) \\
a+\epsilon da \\
b+\epsilon db
\end{array}\right] 
\overset{\text{by (\ref{perturb})}}
\mapsto 
\left[ \begin{array}{c}  f^\mu+\epsilon u\circ f^\mu +o(\epsilon)  \\
a+\epsilon da \\
b+\epsilon db
\end{array}\right] 
\end{align*}
where $u\circ f^\mu(\zeta)=-\frac{1}{\pi}\iint L^\mu\nu(z) R(z,f^\mu(\zeta))dx dy$ and $L^\mu\nu=\left\{ \frac{\nu}{1-|\mu|^2}\frac{\partial f^{\mu}}{\overline{\partial f^{\mu}}} \right\}\circ (f^\mu)^{-1}$. To get the perturbations in the form $f^{\bs \theta+\epsilon d\bs \theta}=f^{\bs\theta}+\epsilon u^{\bs \theta,d\bs \theta}\circ f^{\bs \theta}+o(\epsilon)$ we simplify equation
\begin{align*}
 f^{\bs \theta+\epsilon d\bs \theta}&= (a+\epsilon da) \bigl(f^{\mu} +\epsilon u\circ f^\mu +o(\epsilon)\bigr) + b+\epsilon db\\
 &=f^{\bs \theta}+\epsilon (a u \circ f^{\mu} + da\, f^\mu +db)+o(\epsilon). 
 \end{align*}
Notice
\begin{align*}
 u\circ f^\mu(\zeta)
&=-\frac{1}{\pi}\iint \nu(z) R(f^\mu(z), f^{\mu}(\zeta)) (\partial f^\mu(z))^2 dx dy \\
&=-\frac{1}{\pi}\iint \nu(z) \frac{(b-f^{\bs \theta}(\zeta))(a+b-f^{\bs \theta}(\zeta))}{(f^{\bs\theta}(z)-f^{\bs \theta}(\zeta))(b-f^{\bs\theta}(z))(a+b-f^{\bs\theta}(z))} \frac{ (\partial f^{\bs\theta}(z))^2 }{a}dx dy\\
&=-\frac{1}{a\pi}\iint L^{\bs \theta}\nu(z) \frac{(b-f^{\bs \theta}(\zeta))(a+b-f^{\bs \theta}(\zeta))}{(z-f^{\bs \theta}(\zeta))(b-z)(a+b-z)} dx dy,
\end{align*}
where $L^{\bs \theta}\nu := \left\{ \frac{\nu}{1-|\mu|^2}\frac{\partial f^{\bs\theta}}{\overline{\partial f^{\bs\theta}}} \right\}\circ (f^{\bs\theta})^{-1}$. Therefore, finally, we have
\begin{equation} 
\label{555}
f^{\bs \theta+\epsilon d\bs \theta}=f^{\bs \theta}+\epsilon u^{\bs \theta,d\bs \theta}\circ f^{\bs \theta}+o(\epsilon) 
  \end{equation}
  where
\begin{equation}\label{77}
u^{\bs\theta,d\bs \theta}(\zeta)= db+\frac{da}{a} (\zeta -b) -\frac{1}{\pi} \iint L^{\bs \theta}\nu(z)   \frac{(b-\zeta)(a+b-\zeta)}{(z-\zeta)(b-z)(a+b-z)}  dx dy.  \end{equation}
If we have the nice situation that  $L^{\bs \theta}\nu\in L_p$ for some $p>2$ it can be shown that   $\overline \partial u^{\bs\theta,d\bs\theta} = L^{\bs \theta} \nu$ in the distributional sense where $u^{\bs \theta,d\bs\theta}(b)=db$ and $u^{\bs \theta,d\bs\theta}(a+b)=da+db$.

\section{Rate of Change of the Likelihood}
\label{RateLike}
We now return to the observation scenario where we have {\it iid} observations $X_1,\ldots,
X_n$ from the measure $\Bbb P\circ f^{\bs \theta}$ where $\Bbb P$ is a known measure on $\Bbb R^2$ and $f^{\bs \theta}$ is known only to be a member of the class of invertible transformations $\mathcal F$ given by (\ref{classesofdef}).
We also suppose that $\Bbb P$  has a strictly positive differentiable density with respect to Lebesgue measure on $\Bbb R^2$ and write the density $\exp H$ for some  differentiable function $H\colon\mathbb R^2\rightarrow \mathbb R$.  
 Since $H$ is assumed to be known, the likelihood of the observations $X_1,\ldots,X_n$, as a function of $\bs \theta\in\Theta$, is given by (\ref{loglike}). Let $\bigl|J^{\bs\theta+\epsilon d\bs\theta}\bigr(x)|$ denote the determinant of the Jacobian of the map $f^{\bs \theta+\epsilon d\bs\theta}$ evaluated at $x\in\Bbb R^2$. 
Notice that equations (\ref{555}) and (\ref{jacobiId}) give 
\begin{align}
 \frac{d}{d\epsilon}\Bigl[\log \bigl | J^{\bs \theta+\epsilon d\bs\theta} (x) \bigr |\Bigr]_{\epsilon=0}&= \text{div}\, u^{\bs \theta,d\bs\theta}\circ f^{\bs\theta}(x) \\
\frac{d}{d\epsilon}\Bigl[ H\circ f^{\bs\theta+\epsilon d\bs\theta}(x)\Bigr]_{\epsilon=0}&=\bigl\langle u^{\bs \theta,d\bs\theta}\circ f^{\bs\theta}(x),\nabla H\circ f^{\bs \theta}(x)\bigr\rangle
\end{align}
for all $x\in\Bbb R^2$
where $u^{\bs \theta,d\bs\theta}$ is given by equation (\ref{77}) and $\bigl\langle \cdot,\cdot \bigr\rangle$ denotes dot product in $\Bbb R^2$. Therefore
\begin{align*} 
\dot \ell[\bs \theta](d\bs\theta)&:=\frac{d }{d\epsilon}\Bigl[\ell(\bs \theta+\epsilon d\bs\theta) \Bigr]_{\epsilon=0} \\
 &=\frac{d}{d\epsilon} \left[ \sum_{k=1}^n\log |J^{\bs \theta+\epsilon d\bs\theta}(X_k)| +H\circ f^{\bs \theta+\epsilon d\bs\theta}(X_k)  \right]_{\epsilon=0} \\
 &=\sum_{k=1}^n \text{div}\,u^{\bs\theta, d\bs \theta}(Y_k)+ \bigl \langle  u^{\bs\theta,d\bs\theta}(Y_k) ,\nabla H(Y_k)\bigr\rangle, 
\end{align*}
where $Y_k=f^{\bs\theta}(X_k)$ and $u^{\bs \theta,d\bs\theta}$ is given by (\ref{77}).
Notice that this gives formula (\ref{like}) given in the introduction.
Finally we mention that $\dot\ell[\bs\theta](\cdot)$ is linear over the reals so that
\[ \dot\ell[\bs \theta](d\bs\theta) = \sum_k d\theta_k \dot\ell[\bs \theta](e_k) \]
where  $\bs \theta=\sum_k \theta_k e_k$ and $d\bs \theta=\sum_k d\theta_k e_k$ where where $e_k=(0,\ldots0,1,0,\ldots)$ with the non-zero element at the $k^\text{th}$ index.


\section{Algorithmic details}
\label{algo}
In this section we discuss the algorithmic details for a gradient ascent of the likelihood $\ell(\bs \theta)$. In our implementation of the algorithm, we slightly adjusted these techniques by using quasi-Newton updates instead of gradient ascent updates. However, the essential components of the algorithm are still captured by gradient ascent while allowing succinct exposition. Therefore, we reserve the details of the quasi-Newton updates for the next section.

The gradient ascent will be done by truncating the parameter space $\bs \theta^N:=(\theta_1,\ldots,\theta_N)$ and using (\ref{like}) to write,
\begin{equation}
\label{ggrad}
 \nabla \ell (\bs \theta^N)=  \begin{pmatrix} \dot{\ell}[\bs\theta^N](e_1)\\\vdots\\ \dot{\ell}[\bs\theta^N](e_N) \end{pmatrix} =\sum_{k=1}^n \begin{pmatrix}  \text{div}\, u_1(Y_k) +\langle u_1(Y_k),\nabla H(Y_k)\rangle   \\\vdots\\    \text{div}\, u_N(Y_k) +\langle u_N(Y_k),\nabla H(Y_k)\rangle\end{pmatrix},
 \end{equation}
where $u_1:=u^{\bs \theta^N,e_1},\ldots,u_N=u^{\bs \theta^N,e_N}$ and $Y_k=f(X_k)$.
Gradient ascent is then written as a solution to,
\begin{equation}
\label{gradd}
 \frac{d}{d t} \bs\theta^N_t =\nabla \ell (\bs \theta_t^N)- \lambda \nabla \mathcal J(\bs \theta_t^N),
 \end{equation}
where $\mathcal J$ is a regularization penalty.
 The most difficult part of of this algorithm is solving $u_1,\ldots,u_N$ at each step in a discretization of (\ref{gradd}). We start this section with a detailed discussion of this problem and finish with a few comments on the recursive nature of the gradient ascent.

To find the vector fields $u_1,\ldots,u_N$ in (\ref{ggrad}) one needs compute the singular integral (\ref{77}). When $\nu$ has compact support, the integral (\ref{77}) has the form
$(P L^{\bs\theta}\nu)(\zeta) + (\text{linear function of  $\zeta$})$ where  $P h(\zeta):= -\frac{1}{\pi} \iint h(z)/(z-\zeta)dxdy$ is the Cauchy Transform.  After a simple scaling, one has available the computational techniques found in  \cite{daripa99singular} for computing the Cauchy Transform over the unit disk.
In particular, suppose $h$ is a function with  compact support in $\Omega$ and let $c$ be a constant such that $c\geq |z|$ for all $z\in\Omega$. Then $h_c(z):=h(c z)$  has support in the unit disk $\Bbb D$ and 
$ P h(c\, \zeta)= c Ph_c(\zeta)=-\frac{c}{\pi} \iint_{\Bbb D} h_c(z)/(z-\zeta)dxdy$ for all $\zeta\in\Bbb D$. Now one can use Diripa and Mashat's method to approximate $\iint_{\Bbb D} h_c(z)/(z-\zeta)dxdy$. Finally  one needs to find the linear correction to $P L^{\bs\theta}\nu$ for the integral  (\ref{77}). This is accomplished by the two necessary conditions $u^{\bs\theta,d\bs\theta}(a)=da$ and $u^{\bs\theta,d\bs\theta}(a+b)=da+db$.

A second approach for approximating $u^{\bs \theta,d\bs \theta}$ is to  solve the nonhomogeneous Cauchy-Riemann equations $\overline \partial u^{\bs \theta,d\bs\theta}= L^{\bs \theta} \nu$. The solution is only unique up to additive analytic functions, however, if it is known that $L^{\bs \theta}\nu \in C^2_0$ then $P L^{\bs \theta}\nu\in C^2$  and $\partial P L^{\bs \theta}\nu \in L_2$ (see Lemma 2, page 52 of \cite{ahlfors2}). Therefore any continuous solution to $\overline \partial u^{\bs\theta,d\bs\theta}= L^{\bs \theta} \nu$ which satisfies $\partial u^{\bs\theta,d\bs\theta}\in L_2$ is an additive constant difference of $PL^{\bs \theta}\nu$. 
Therefore one can find  $u^{\bs\theta,d\bs \theta}$ by  solving the nonhomogeneous Cauchy-Riemann equation $\overline \partial u^{\bs\theta,d\bs\theta}= L^{\bs \theta} \nu$  with free boundary conditions on $\partial \Omega$, where $\Omega$ is the support of $ L^{\bs \theta} \nu$.

In our implementation of the algorithm the Diripa and Mashat's method suffered from some instabilities. We instead used an ad-hoc method for trying to invert the equation  $\overline \partial u^{d\bs\theta}= L^{\bs \theta} \nu$ by representing $u^{\bs\theta,d\bs \theta}$ by a latent complex function $w$ such that  $u=\partial w$ (this is basically a stream function and potential function representation of a vector field with the  coordinates switched) then using the fact that $4 \partial\overline\partial=\Delta $, the Lapacian operator, we used a fast Poisson solver in {\sc Matlab} to invert $\Delta w= 4L^{\bs \theta} \nu $. One drawback to this technique is that it potentially differs from $PL^{\bs \theta} \nu$  by an additive analytic function.  In the cases we studied the magnitude of the analytic difference was small enough to be ignored. However, we would prefer Daripa  and Mashat's method but leave it's proper implementation to more adept numerical programmers.

One particularly attractive feature of the algorithm is its recursive nature. 
Suppose at step $m$ of a discretized gradient ascent one has $f^{\bs \theta_m}$ and $f^{\bs\theta_m}_z$ evaluated on a dense grid. We show how to find $f^{\bs \theta_{m+1}}$ and $f^{\bs\theta_{m+1}}_z$  at step $m+1$. 
To makes the following formulas more readable let $\mu^{\bs \theta_m}=\frac{\varphi^{\bs \theta_m}}{1+|\varphi^{\bs\theta_m}|}$ be the complex dilatation of $f^{\bs \theta_m}$, where $\varphi^{\bs \theta_m}=\sum_{k\geq 4} (\bs \theta_m)_k \varphi_k$ is given by the basis expansion in (\ref{classes}).
The gradient update is  $\bs \theta_{m+1}=\bs \theta_m+\epsilon d\bs\theta_m$ where
\[ d\bs \theta_m = (\dot{\mathcal J}[\bs\theta_m](e_1),\dot{\mathcal J}[\bs\theta_m](e_2),\ldots ). \]
Remember that for each $j\geq 4$, 
\[\dot{\mathcal J}[\bs\theta_m](e_j)= \sum_{k=1}^n \text{div}\, u_j(Y_k) +\langle u_j(Y_k),\nabla H(Y_k)\rangle,\] 
where, $Y_k=f^{\bs \theta_m}(X_k)$, $ \overline \partial u_j =L^{\bs \theta_m} \nu_j$ and $\nu_j$ is given by
\begin{align*}
 \nu_j&= \varphi_j\, \frac{2+|\varphi^{\bs \theta_m}| }{2(1+|\varphi^{\bs \theta_m}|)^2} -  \overline {\varphi_j} \,\frac{ (\varphi^{\bs \theta_m})^2}{2|\varphi^{\bs \theta_m}|(1+|\varphi^{\bs \theta_m}|)^2}. 
 \end{align*}
Notice that computing $L^{\bs \theta_m} \nu_j$ can be accomplished by simply deforming the graph of $\frac{\nu_j}{1-|\mu^{\bs \theta_m}|^2} \frac{f_z^{\bs\theta_m}}{\overline{f_z^{\bs\theta_m}}}$ by $f^{\bs\theta_m}$.
Now $f^{\bs \theta_{m+1}}:=f^{\bs \theta_m}+ \epsilon u^{\bs\theta_m,d\bs \theta_m}\circ f^{\bs \theta_m}$ and using the composition rule for $\partial$ and fact that $\overline \partial u^{\bs\theta_m,d\bs \theta_m}=L^{\bs \theta_m}\nu$ we get the following update for $f_z^{\bs\theta_{m+1}}$
\[f_z^{\bs \theta_{m+1}}= f_z^{\bs \theta_m} \left(1+\epsilon\, u_z^{\bs\theta_m,d\bs\theta_m} \circ f^{\bs\theta_m}  + \frac{\epsilon\, \nu\,\overline{\mu^{\bs \theta_m}}}{1-|\mu^{\bs \theta_m}|^2} \right)\]
where $\nu=\sum_{k\geq 4} (d\bs\theta_m)_k \varphi_k$.

%

\begin{figure}
\label{f:showcase}
\centering
\includegraphics[width=5.5 in]{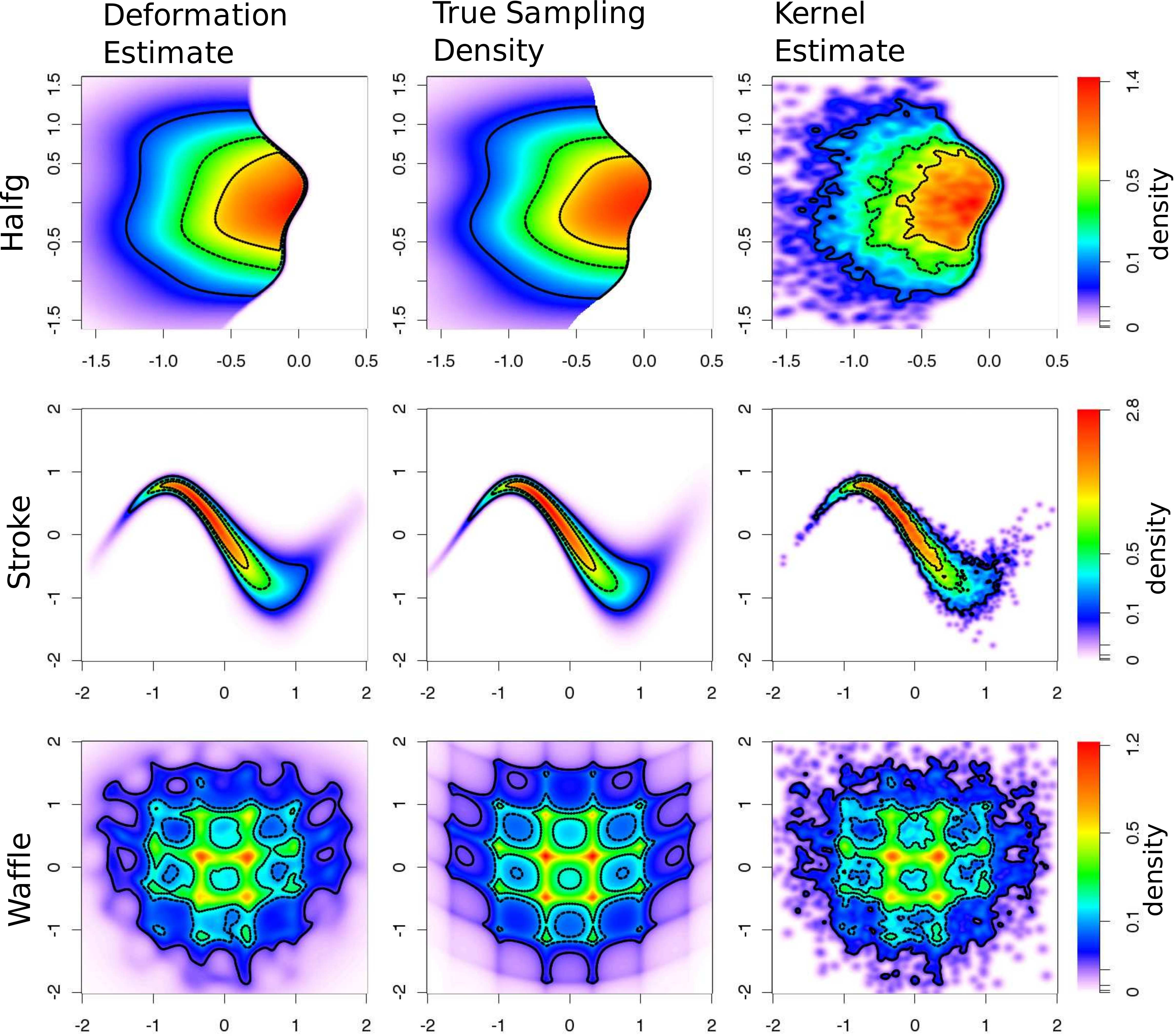}
\caption{Panel of Density Estimates: The true sampling density is shown in the central column; in the left and right columns are density estimates based on the same sample of 10,000 points using respectively the deformation-based methods of this paper and kernel density estimates. There is one row per example: halfg, stroke, waffle. As indicated by the colorbar, the scale for the heatmaps is non-linear to accentuate differences in low-density regions (a square-root transform is used, consistent with Hellinger distance). A solid line is drawn at the contours of the density that capture $95\%$ of the mass; the $50\%$ line is dashed, and the $25\%$ line is dash-dotted.
}
\end{figure}

\section{Simulation Experiments}
\label{empir}

To investigate the quality of the density estimates obtained by the deformation method we performed a series of simulation experiments. We first give an overview of the results; details are explained afterward. Data sets of size 100, 1000,  and 10,000 were generated independently from three distributions that we call halfg, stroke, and waffle. Each example is intended to demonstrate different properties of our estimator; the examples, and the results are illustrated in Figure~1.

The halfg example was chosen to highlight the ability of the deformation method to 1) adapt to sharp edges of the density 2) utilize qualitative knowledge about the form of the density, namely through the use of a softened half-normal as the target density. The stroke example, in which a normal density is stretched out into an ``S'' with the thickness varying along the length, is intended to test the ability of the deformation to capture the overall structure well enough to extend mass into the tails appropriately. The waffle example tests the estimates performance in a complex, bumpy case. Both the stroke and the waffle examples use a radially symmetric bivariate Gaussian distribution with mean 0 as the target. For stroke, the target has standard deviation 0.05; for waffle, the target has standard deviation 0.5. From these choices we find that the method can successfully reconstruct the long thin stroke density by ``stretching'' a small target, and can also match the complex waffle pattern onto a moderate target.

For each data set, the optimal penalty parameter $\lambda$ was chosen to minimize the integrated square error (ISE) difference between the density estimated for that $\lambda$ and the true sampling density. The minimal ISE was recorded and compared with the minimum ISE achieved by kernel density estimation at the optimal bandwidth. To determine the sampling variability of the findings, this was repeated six times for each example. These results, as well as results for Hellinger and $L^1$ distance measures, are tabulated in Tables~1,~2, and~3.
Examples of the estimates achieved on data sets of 10,000 are shown in Figure~1.

\begin{table}
\label{t:results}
\caption{Table of the average integrated squared errors, summarized across six independent realizations. Averages (and parenthetical standard deviations) are reported times $10^4$}
\centering
\begin{tabular}{llrrr}
         &      &  \multicolumn{3}{c}{Sample Size} \\
Case          &  Method    & $10,000$ & $1,000$ & $100$ \\
\hline
Halfg &deformation & 59 ~(4) &  109 (19) & 325 ~(83) \\
       & kernel & 190 ~(8) & 409 (34) & 880 (181) \\
\\
Stroke & deformation & 24 ~(8) & 122 (30) & 909 (209) \\
          & kernel & 157 (18) & 642 (81) & 2354 (454) \\
\\
Waffle & deformation & 64 ~(5) & 222 (30) & 674 ~(30) \\
         & kernel & 74 ~(5) & 259 (26) & 662 ~(49) \\
\end{tabular}
\end{table}

\begin{table}
\label{t:results_Hel}
\caption{Table of the average Hellinger-distance errors, summarized across six independent realizations. Averages (and parenthetical standard deviations) are reported times $10^4$}
\centering
\begin{tabular}{llrrr}
         &      &  \multicolumn{3}{c}{Sample Size} \\
Case          &  Method    & $10,000$ & $1,000$ & $100$ \\
\hline
Halfg &deformation & 863 (17) &  1160 ~(43) & 1915 (250) \\
       & kernel & 1881 (19) & 2718 ~(54) & 3860 (236) \\
\\
Stroke & deformation & 443 (33) & 992 (110) & 2601 (187) \\
          & kernel & 1083 (32) & 2048 ~(53) & 3800 (285) \\
\\
Waffle & deformation & 1021 (31) & 1826 ~(68) & 3002 (116) \\
         & kernel & 1193 (24) & 2105 ~(66) & 3149 ~(78) \\
\end{tabular}
\end{table}

\begin{table}
\label{t:results_L1}
\caption{Table of the average L1-distance errors, summarized across six independent realizations. Averages (and parenthetical standard deviations) are reported times $10^4$}
\centering
\begin{tabular}{llrrr}
         &      &  \multicolumn{3}{c}{Sample Size} \\
Case          &  Method    & $10,000$ & $1,000$ & $100$ \\
\hline
Halfg &deformation & 544 (46) &  955 ~(93) & 1808 (250) \\
       & kernel & 1385 (35) & 2382 (115) & 3980 (475) \\
&&&&\\
Stroke & deformation & 487 (71) & 1124 (120) & 3104 (259) \\
          & kernel & 1268 (53) & 2572 (124) & 5018 (442) \\
&&&&\\
Waffle & deformation & 1535 (48) & 2813 (136) & 4814 (139) \\
         & kernel & 1726 (38) & 3153 (113) & 4969 (117) \\
\end{tabular}
\end{table}

\subsection{Sampling Densities}\label{s:sampling} In this subsection we give the detailed recipes for the sampling distributions and target densities used in the examples.

\paragraph{Halfg:} Let $x=(x_1,x_2)^t$, let $\Psi(x)=(x_1+h(x_2),x_2)^t$, where $h(u)=\sin(5 u)/15-u\tanh(u)/3$. The data were generated by sampling from a symmetric bivariate normal with standard deviation $1/2$, rejecting points in the right half plane, and applying the transformation $\Psi$ to the remaining points. In effect, then, the density of the observed points is a deformed half-normal. Notice that the indicator introduces a smoothly curved sharp edge to the density:
\[
2 (1/\sqrt{2 \pi})^2(1/2)^2 \exp(-1/8( (x_1-h(x_2))^2 + x_2^2)\mathbf{1}_{x_1-h(x_2)\le 0}
\]

For the halfg example, instead of using a standard Gaussian as our target we used a softened half-normal; the softening was done so that the target would be a continuously differentiable density to facilitate optimization, yet still describe a half-normal-like distribution. Specifically, the target distribution is described by independently drawing $X_2$ from a normal with mean 0 and standard deviation $\frac{1}{2}$, and $X_1$ from the mixture of half-normals with density $\gamma [2\phi_{\sigma_+}(x){\bf I}_{x \ge 0}] + (1-\gamma) [2\phi_{\sigma_{-}}(x) {\bf I}_{x<0}]$, where $\sigma_{-}=\frac{1}{2}$, $\sigma_{+}=\frac{1}{50}$, $\gamma=(1+\sigma_{-}/\sigma_{+})^{-1}$, and $\phi_\sigma$ is the density of a normal with mean 0 and standard deviation $\sigma$.

\paragraph{Stroke:} Let $\Psi(x)=\left(x_1 /2,  \frac{1}{20}(3+2 \tanh(x_1)) x_2 - \frac{4}{5}\sin(x_1)\right)^t$. The samples are generated by sampling $X=(X_1, X_2)^t$ as a pair of independent standard normals, and applying transformation $\Psi$. The target density was taken to be a mean $0$ symmetric bivariate normal with standard-deviation $0.05$.

\paragraph{Waffle:} Let $\Psi(x)=\frac{2}{3} \left( x_1+ \sin(2 \pi x_1)/ (2 \pi s), x_2+\cos(2 \pi x_2)/(2 \pi s) + (x_1/3)^2 \right)^t$, where we take $s=\frac{1}{2}$. The samples are generated by sampling $X=(X_1, X_2)^t$ as a pair of independent standard normals, and applying transformation $\Psi$. The target density was taken to be a mean $0$ symmetric bivariate normal with standard-deviation $0.5$.

\subsection{Basis Functions $\varphi_k$ and Penalty $\mathcal J(\bs \theta)$.}
Here we define the basis functions $ \varphi_k$, in (\ref{classes}), and the regularization penalty $\mathcal J$, in (\ref{gradd}), used for our simulations.  To motivate our choice of basis functions $\varphi_k$,  we note  the  Fourier representation of $L_2$ functions on $[-L,L]^2$:
\begin{align*} 
\varphi(x+iy)&=\sum_{k_1,k_2\in \Bbb Z}  (a_{k_1,k_2}+i b_{k_1,k_2}) e^{i 2\pi (k_1x+k_2y)/2L} \\
&=\sum_{k_1,k_2\in \Bbb Z}  a_{k_1,k_2}e^{i 2\pi (k_1x+k_2y)/2L} +\sum_{k_1,k_2\in \Bbb Z} b_{k_1,k_2} ie^{i 2\pi (k_1x+k_2y)/2L},
\end{align*}
for real coefficients $a_{k_1,k_2}, b_{k_1,k_2}$. These Fourier basis functions are smoothly truncated to zero outside of the disk of radius $L$ by multiplying each basis function by  $\bs T(x,y)=[1-((x/L)^2+(y/L)^2)^4]\bs 1_{\{x^2+y^2<L^2\}}(x,y)$ where $\bs 1_{\{x^2+y^2<L^2\}}$ is the indicator of the  disk of radius $L$. The infinite series is then truncated to a finite expansion to get the following $(2N+1)^2$ basis elements used in our simulations to define $\{ \varphi_k\}$:
\[ \{ \varphi_k \}_{1\leq k\leq (2N+1)^2}:=\bigcup_{k_1,k_2=-N}^N\left\{e^{i 2\pi (k_1x+k_2y)/2L} \bs T(x,y),  ie^{i 2\pi (k_1x+k_2y)/2L} \bs T(x,y)\right\}.\] 

The regularization penalty $\mathcal J$ in (\ref{gradd}) is defined by, 
\[\mathcal J( \bs \theta):= \sum_{k_1,k_2} (k_1^2+k_2^2)^2 (a_{k_1,k_2}^2 +b^2_{k_1,k_2} ), \]
where $\bs \theta$ is the parameter vector composed of  the basis coefficients $a_{k_1,k_2}$, $b_{k_1,k_2}$ and the scale terms $a_1,a_2,b_1,b_2$ defined in Section (\ref{classes}).  Our choice of the penalty is motivated by the Fourier representation of the spline penalty $\int_{[-L,L]^2} |\Delta \varphi(x,y)|^2dxdy\propto \sum_{\bs k\in \Bbb Z^2}  |\bs k|^4|\widehat \varphi(\bs k)|^2$.

All of our simulations recorded in Tables~1,~2, and~3
used $N=6$ except that for the datasets of size 10,000 we used $N=10$ for the waffle and $N=8$ for the stroke in the hope that this would allow us to capture finer detail. Generally the results do not appear to be sensitive to the choice of $N$. For example, some of the waffle/stroke examples were also attempted with $N=6$; aside from one exceptional case, the ISE's obtained differ by no more than $4 \times 10^{-4}$ from the values reported in the table. In the exceptional case, the optimizer failed to improve upon the initial conditions; upon restarting with slightly different initial values it appeared to behave normally. The small dependence on $N$ can be attributed to the high coefficients of the penalty on high frequency terms.

\subsection{Optimization Details}
For each data set, a grid of values of $\lambda$ was used, namely: $10^{1.5}, 10^{1.25}, 10^{1}, \dots, 10^{-2.75}, 10^{-3}$.
The optimization was performed for each $\lambda$ in turn, from the largest to the smallest, so that the parameter estimates and estimates of the local Hessian of the objective (as produced by the optimizer) from the more highly regularized cases could serve as initial values for the less regularized problems.

The actual optimization was performed by an adaptation of the BFGS algorithm (see \cite{langeOpt}). The coefficients of the basis for the dilatation were broken into real and complex parts and the gradient was expressed as a real-valued vector with respect to this partition. The Hessian estimate was initialized as a diagonal matrix with 1000 on the diagonal. All steps proposed by BFGS were restricted to have a maximum length of 0.02. Very long steps are to be avoided because they may introduce numerical error. After every successful step, the Hessian-estimate was updated in the customary BFGS manner.

If a proposed step is too long, the step is shrunk using the trust-region technique (see \cite{langeOpt}); i.e. essentially by temporarily increasing the diagonal of the local Hessian in the Newton-type step. The end result of this technique is that each step taken is optimal, according to  the current quadratic approximation to the objective function, subject to the constraint that it lie in a ball of the given radius. Similarly, if the step turned out not to reduce the objective, the step was retracted and then the target step length was halved. If ten successive halvings fail to result in a downward step, the failure was noted and the current parameter estimates were stored and the associated deformation was recomputed from scratch. This recomputation also occurs on every tenth step to reduce the accumulation of numerical errors. If descent fails again at this point, the optimization is terminated. 

%
\section{Discussion}
The simulation results displayed in Tables 
1, 2, and 3, as well as 
Figure
1
illustrate the potential of using transformations for density estimation. We believe that by penalizing the transformation of the distribution to be smooth, rather than the directly forcing the density estimate to be smooth, we introduce an interesting new regularization strategy. In the figure, we see that this approach produces qualitatively pleasingly smooth estimates of the tails. In the case of the halfg example, we see that it successfully incorporated the prior information that a smooth edge is present, resulting in an efficient estimate of the location and shape of this curve. 

The quantitative improvements is most dramatic for the  stroke example in which the relative improvement in mean integrated squared error is respectively $85\%$, $81\%$, and $61\%$ for sample sizes of $10,000$, $1,000$, and $100$. These results sound only slightly less impressive if reported in terms of the $L^2$-norm, instead of the ISE (its square), namely improvements of $61\%$, $56\%$, and $38\%$ respectively. These percentages roughly agree with the $\%$-improvements seen in terms of the Hellinger and $L^1$ metrics. The results for the halfg example are also good, with improvements ranging from $39\%$ to $69\%$ across all sample sizes and criteria. For example, even at the sample size of $100$, use of the prior information that the sampling density is approximately a perturbed half-normal, results in an improvement of $55\%$ over kernel density estimation in terms of mean Hellinger distance. The results are least impressive for the waffle example, whose complex, bumpy structure is well suited to the kernel-estimate. The relative improvements range from $7\%$ to $14\%$ (depending on the metric used) at samples sizes of $10,000$ or $1,000$. At the smallest sample size of $100$, the mean ISE is worsened by $2\%$ (not significant by nominal $.05$-level t-test), although on these same six realizations, there is a (nominally significant) $3\%$ to $5\%$ improvement in the $L^1$ and Hellinger metrics.

 The contributions of this paper include the development of a flexible class of two dimensional transformations and a derivation of the rate of change of the log-likelihood as it depends on the unknown transformation. These variational results are then used to construct a gradient-based algorithm for estimating the transformation that maximizes a penalized log-likelihood. In Section \ref{algo} we discuss how to approximate the vector field perturbation of $f^{\bs \theta}$ by solving a  differential equation of the form $\overline \partial u^{\bs \theta,d\bs \theta}= L^{\bs \theta}\nu $. Our results are in two dimensions, however, it is our hope that using transformations for density estimation can ultimately be useful in high dimensional density estimation by smoothly interpolating density in sparse areas while still preserving \lq sharp' features of the true density.
 We consider this  a first step toward both numerical and theoretical exploration of using invertible maps in non-parametric and semi-parametric density estimation in general dimensions. 

\bibliography{references}

\end{document}